\documentclass[11pt,twoside]{article}
\usepackage{asp2010}

\resetcounters

\begin{document}

\title{Period -- mass-loss rate relation of Miras with and without Tc}
\author{Stefan Uttenthaler$^1$
  \affil{$^1$University of Vienna, Department of Astrophysics,
    T\"urkenschanzstra\ss e 17, 1180 Vienna, Austria}
}

\begin{abstract}
We present the discovery that Mira variables separate in two distinct sequences
in a near- to mid-IR color versus pulsation period diagram, if a distinction is
made with respect to the presence of technetium (Tc) in the stars. Tc is an
indicator of recent or ongoing deep mixing during a third dredge-up event. At a
given period, the Tc-poor Miras are redder in $K-[22]$ (i.e.\ have higher dust
mass-loss rate) than the Tc-rich Miras. This is counter-intuitive since the
Tc-rich Miras are expected to be more evolved and should have a higher mass-loss
rate. In this contribution we give an update on this recently discovered
conundrum.
\end{abstract}

\section{The phenomenon}

Figure~\ref{K22_P} presents the phenomenon discovered by \citet{Utt_2013}. The
figure shows a diagram of the $K-[22]$ color of Miras versus their pulsation
period. [22] is the magnitude in the 22\,$\mu$m band of the WISE space
observatory, and the near- to mid-infrared color $K-[22]$ is an indicator of
the dust mass-loss rate ($\dot{M}_{\rm dust}$) of the stars. It is related to the
total mass-loss rate via the gas-to-dust ratio $\delta$. Stars in
Fig.~\ref{K22_P} are marked according to their Tc content. Tc is an element with
only radioactive isotopes that is produced by the s-process in the deep interior
of AGB stars. It is an indicator of recent or ongoing third dredge-up (3DUP), a
deep mixing event taking place in AGB stars.

\articlefigure[width=\linewidth,bb=93 368 548 700, clip]{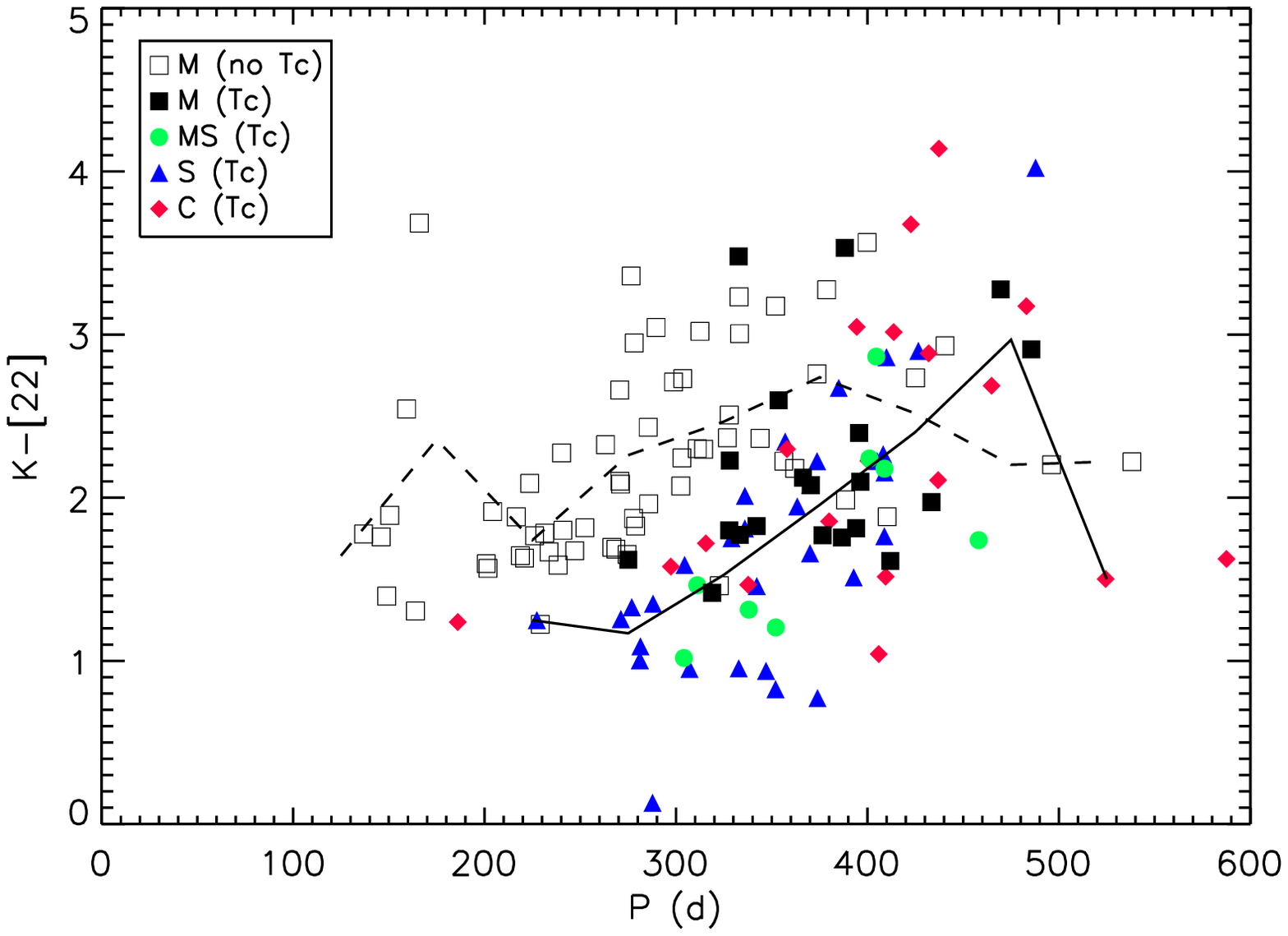}{K22_P}{Mira stars of different atmospheric chemistry (spectral type) in the $K-[22]$ vs.\ $P$ diagram. Empty symbols are for Tc-poor, filled symbols are for Tc-rich stars. The dashed line indicates the run of the mean $K-[22]$ color in 50\,d period bins for the Tc-poor stars, while the solid line shows this run for the Tc-rich stars. Taken from \citet{Utt_2013}.}

The diagram in Fig.~\ref{K22_P} clearly shows that the Tc-poor (open symbols)
and the Tc-rich (filled) Miras occupy different regions, each of the two groups
forming a sequence of increasing $K-[22]$ with increasing period. The
counter-intuitive aspect is that at a given period, the Tc-poor Miras have a
{\it redder} $K-[22]$ color than the Tc-rich ones, indicating a higher dust
mass-loss rate. This is not what one would na\"{i}vely expect because the
Tc-rich Miras are more evolved than the Tc-poor ones, a phase when also the
(dust) mass-loss rate is expected to be higher. This figure demonstrates that
this simple picture must be wrong. Note that this phenomenon is not inherent to
the [22] band; mid-IR bands ranging from AKARI [9] to IRAS [60] were
investigated, too, all of which show the same two sequences.

There are a few objects that blur the otherwise clear separation between
Tc-poor and Tc-rich Miras in that diagram. There are two Tc-rich, M-type Miras
(filled black squares) at $P\approx330$\,d and $\approx390$\,d and
$K-[22]\approx3.5$. These are o~Cet and R~Aqr, both of which are well-known
binary AGB stars. The reddest S-type Mira (blue triangle), at $P\approx490$\,
and $K-[22]\approx4.0$, is W~Aql, also a binary AGB star. It seems that binary
AGB stars are particularly red for their period. Possibly, the presence of a
companion to an extended red giant star enhances either the dust production in
the system or the (dust) mass-loss rate from the system. There is a very red,
Tc-poor Mira (open square) at $P\approx165$\,d, which is R~Cet. We tentatively
predict that this is a binary star, too.

On the other hand, there are Tc-poor Miras at long period, but relatively blue
$K-[22]$ color. The Tc-poor Miras with the longest pulsation period in the
sample are R~Nor and R~Cen, both of which have been suggested to be massive
AGB stars ($M\gtrsim4M_{\odot}$). The dominant neutron source in these stars is
from the $\alpha$-capture on $^{22}$Ne, which is predicted to produce only very
little Tc \citep{Gar_etal_2013}. A similar case could be W~Hya, at
$P\approx390$\,d and $K-[22]\approx2.0$, which has also been speculated to be of
relatively high mass.

We also investigated the distribution of semi-regular variables (SRVs) in that
diagram, in particular the Tc-rich ones. A sequence of increasing mass-loss
rate with increasing period is formed by mainly S- and C-type SRVs at short
periods ($0<P\lesssim200$\,d) and rather blue colors ($K-[22]\lesssim1.2$). It
is possible that some of the SRVs at the top of this sequence switch to overtone
pulsation, i.e.\ to the Mira phase, and continue their evolution there. However,
this seems to be possible only for the S- and C-type stars, not for the M-type
stars.

Also, there are a few SRVs with much redder $K-[22]$ color than the mentioned
sequence. Some of these stars are suspected or known binaries
(e.g.\ $\pi^{1}$~Gru, L$^{2}$~Pup), which leads one to suspect that also in this
group the binarity has a considerable effect on the (dust) mass-loss rate.

\section{ISO dust spectra}

It is instructive to also inspect the dust spectra of Miras with and without Tc.
Spectra of AGB stars in the mid-IR have been observed by ISO and processed and
published by \citet{Slo_2003a}. Unfortunately, there is not much overlap between
the ISO sample and the sample of AGB stars with known Tc content discussed here.
Nevertheless, already the few stars in common show interesting things.
Figure~\ref{ISO} shows a comparison of the ISO spectra of the Tc-poor Mira
RR~Aql and the Tc-rich star R~Hya. As shown in the figure legend, the two Miras
have very similar pulsation periods, but quite different $K-[22]$ colors.

\articlefigure[width=10cm,bb=82 370 538 698, clip]{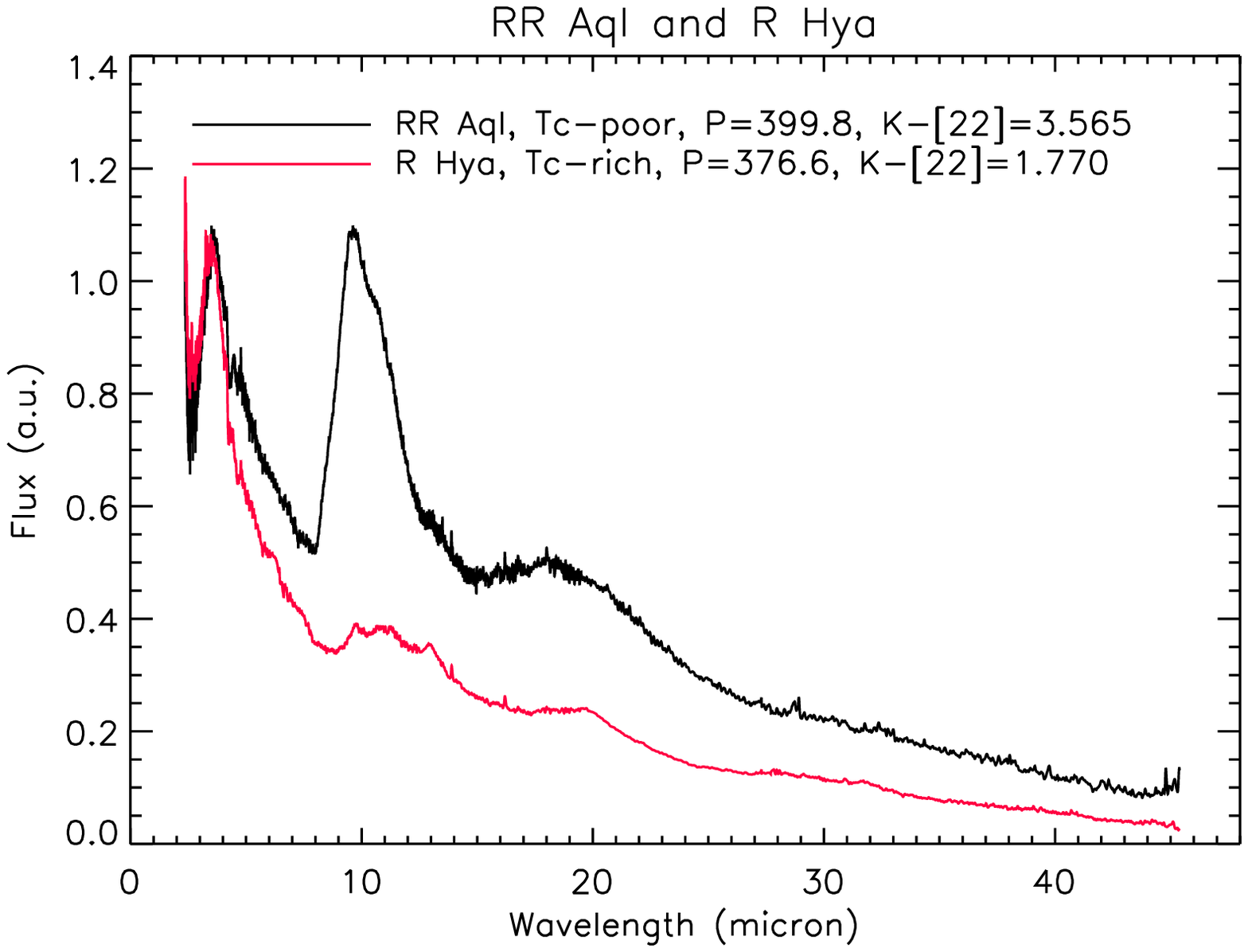}{ISO}{ISO dust spectra of the stars RR~Aql (Tc-poor, black line) and R~Hya (Tc-rich, red line). The spectra are normalised to a mean of 1.0 in the range between 3 and 4\,$\mu$m.}

The spectra of the Tc-rich Miras S~Vir and T~Cep, also with very similar
pulsation periods, look very similar to R~Hya. Clearly, the Tc-poor and Tc-rich
Miras are distinctly different also in the ISO dust spectra, with the Tc-rich
stars having a much lower mid-IR dust excess than the Tc-poor ones over a wide
wavelength range.

\citet{Slo_2003b} investigated the appearance of the 13\,$\mu$m feature in the
ISO sample and its correlation to other dust features and stellar parameters.
This feature is also visible in the spectra of some of the Tc-rich Miras, but a
clear correlation is not obvious in the small overlap sample. Nevertheless,
within each of the groups ``Tc-poor Miras'', ``Tc-rich Miras'', and
``Tc-rich SRVs'', the sources showing the 13\,$\mu$m feature are the bluest
ones, and those without the 13\,$\mu$m feature are the reddest ones in $K-[22]$
(there is only one Tc-poor SRV with an ISO spectrum). This lends support to the
conclusion of \citet{Slo_2003b} that the 13\,$\mu$m feature is stronger in
objects with lower IR excess and thus lower mass-loss rates. Hence, the
13\,$\mu$m feature in Tc-rich Miras is probably not a result of the 3DUP, but a
result of the low mass-loss rate (which could be low because of 3DUP, see
below). However, due to the small sample of AGB stars with observed dust
spectrum and information on the Tc content, one must be cautious with strong
conclusions.

\section{Hypotheses for explanation}

How can we interpret the two sequences? How do stars evolve in the $K-[22]$
vs.\ $P$ diagram? In \citet{Utt_2013}, two main hypotheses were put forward.

\begin{itemize}
\item A Tc-poor Mira switches to the Tc-rich sequence upon a 3DUP event.
\item Tc-poor and Tc-rich Miras are groups of different mass, i.e.\ the
  Tc-rich ones are more massive than the Tc-poor ones, and the two groups
  simply evolve differently in the diagram.
\end{itemize}

The first hypothesis offers two options: i) the pulsation period increases by
the dredge-up of material (e.g.\ due to changed molecular opacities due to
changed elemental abundances), and ii) the mass-loss properties
($\dot{M}_{\rm dust}$, gas-to-dust ratio $\delta$, expansion velocity
$v_{\rm exp}$) change such that the $K-[22]$ color becomes bluer. A recent paper
by \citet{Scholz_2014} investigated the reaction of pulsation models on a
moderate abundance change, e.g.\ due to 3DUP. They found that the pulsation
period stays virtually unchanged by moderate abundance changes. If this is
correct, then the first option of hypothesis 1 is excluded.

In \citet{Utt_2013}, hypothesis 2 was favored as the correct explanation for
the phenomenon. However, this conclusion was based on a premature analysis of
the mean distance from the Galactic mid-plane of the two groups. The mean
distance from the plane is an indicator of the mean mass of a class of stars,
lower mean distance means higher average mass. It is known that long-period
Miras are more massive than short-period Miras because on average they are
closer to the Galactic mid-plane. The Tc-rich Miras are on average closer to the
Galactic mid-plane than the Tc-poor Miras, suggesting that they have a higher
mass. However, this is biased by the fact that Tc-rich Miras simply have longer
periods on average. A closer inspection shows that in 50\,d bins in the period
range between 200 and 450\,d, where the two groups overlap, both groups have
essentially the same average distance from the plane! Hence, at a given period,
Tc-poor and Tc-rich Miras have the same average mass.

Another indicator of the stellar mass could be the pulsation amplitude. More
massive stars are expected to have a smaller pulsation amplitude. For our sample
we inspected the amplitude in the V-band. We find that there is no significant
difference in the pulsation amplitude of Tc-poor and Tc-rich Miras of spectral
type M, MS, and S. Only the C-type Miras have clearly lower amplitudes because
the molecules that have absorption bands in the visual range are much less
temperature-sensitive in C-rich than in O-rich atmospheres. These pieces of
evidence lead us to believe that hypothesis 2 is probably wrong, and that the
second option offered by hypothesis 1, change in mass-loss properties by 3DUP,
appears most likely to be the correct explanation for our observations.

Are there possible physical mechanisms with which we can understand a decrease
of the (dust) mass-loss rate of a Mira upon 3DUP? It was suggested by
\citet{Hoef_2008} that scattering of photons off silicate grains of size
$\sim1\mu$m could be the driving mechanism of mass loss from oxygen-rich, M-type
giants. This mechanism is successful in explaining the photometry of M-type
giants \citep{Bladh_2013}. We suggest here that the radioactivity of the
material mixed to the surface during a 3DUP event could cause these large dust
grains to dissolve again and thus decrease the mass-loss rate. Radioactively
unstable isotopes may be integrated in the large dust grains, where they then
decay. If the decay energy is deposited in the grain, it may be heated so much
that it gets (partially) destroyed. Even a partial destruction could reduce the
mass-loss rate already substantially because the scattering is most efficient
at wavelengths of $\sim1\mu$m where the maximum of the stellar flux is reached.
Detailed calculations would be required to see if this hitherto unaccounted
effect could be at work in the Tc-rich Miras.

\section{Application to bulge AGB stars}

The separation of Tc-poor and Tc-rich Miras in Fig.~\ref{K22_P} is so clear that
it may actually be used to distinguish stars that did undergo a 3DUP event from
those that did not. A line drawn between the points $(P,K-[22])=(120,0)$ and
$(520,4.2)$ separates the two sequences quite well: 85.4\% of the Tc-poor Miras
are above this line, while 87.2\% of the Tc-rich ones are below, even without
excluding the binary stars or any other special cases. This may be used to
investigate the occurrence of 3DUP in stellar populations other than the solar
neighborhood. Data were collected from Galactic bulge Miras via the MACHO
survey (pulsation period), as well as photometry from 2MASS and WISE. The
photometry was dereddened. In this data set of 1091 Miras, 82.9\% of the stars
are above the line defined above. This would mean that only a small number of
Galactic bulge Miras undergoes 3DUP, consistent with the lack of C-stars in the
bulge.

\section{Conclusions}

We conclude that 3DUP appears to be doing something to the mass-loss properties
of AGB variables, though we do not yet know how this is happening. This
conundrum might teach us an important lesson about the mass-loss mechanism in
O-rich Miras, which we better try to understand.
Unfortunately, the author is a spare-time astronomer now, which limits the
possibilities to investigate this phenomenon. Any support by interested
researchers is warmly welcome!

\acknowledgements We thank Klaus Bernhard (Linz, Austria) for collecting and
providing the MACHO data for the Galactic bulge Miras.

\bibliography{uttenthaler}

\end{document}